# Structural and electronic properties of Al nanowires: an ab initio pseudopotential study

Jin-Cheng Zheng[a)*], Hui-Qiong Wang[b)], A. T. S. Wee and C. H. A. Huan

Department of Physics, National University of Singapore, Lower Kent Ridge Road, Singapore 119260.
* **Corresponding author.** Email: jincheng_zheng@yahoo.com  (J-C Zheng)

**Abstract**

The stability and electronic structure of a single monatomic Al wire has been studied using the *ab initio* pseudopotential method. The Al wire undergoes two structural rearrangements under compression, i.e., zigzag configurations at angles of 140º and 60º. The evolution of electronic structures of the Al chain as a function of structural phase transition has been investigated. The relationship between electronic structure and geometric stability is also discussed. The $2p$ bands in the Al nanowire are shown to play a critical role in its stability. The effects of density functionals (GGA and LDA) on cohesive energy and bond length of Al nanostructures (dimmer, chains, and monolayers) are also examined. The link between low dimensional 0D structure (dimmer) to high dimensional 3D bulk Al is estimated. An example of optimized tip-suspended finite atomic chain is presented to bridge the gap between hypothetical infinite chains and experimental finite chains.

**1. Introduction**

Low-dimensional systems such as mono-atomic nanowires and clusters are of interest both from a fundamental point of view and for potential nano-device applications. Recently, monoatomic chains of gold atoms have been fabricated [1,2], and metallic nanowire contacts can be obtained by scanning tunneling microscopy (STM) [3-5] and mechanically controllable break junctions (MCB) [6]. Several experimental studies such as measurements of conductance and applied force [7], and theoretical calculations such as atomistic [8,9], continuous [10,11], mixed [12,13] model simulations and first-principles calculations [14-19] have been performed to understand the conduction, geometric, mechanical, and electronic properties of these metal wires, especially gold nanowires. In a recent example, Sánchez-Portal *et al.* [18] predicted the spinning zigzag shape of monatomic gold wires, and explained current transmission electron microscopy (TEM) results [1]. Relatively few publications have focused on other metal nanowires. The expectation that Al nanowires should also have unique properties is the motivation of this work. Rubio et al [20] took the first step to simulate the one-dimensional chains of Al atoms in BN nanotubes and showed interaction effects. Using an eigenchannel decomposition (ECD) combined with the first-principles recursion-transfer matrix (RTM) method, Kobayashi et al [21] calculated the electron transport through the three-Al-atom wire between jellium electrodes. More recently, Sen et al[19] studied the structure of Al nanowires from first-principles calculations. The electronic structures of metal nanowires under different constraint conditions (e.g., compression constraint) appear to play a critical role on the electron transport properties of STM-suspended nanowires. In this work, we perform first principles calculations to address the questions of structure and stability, as well as the evolution of electronic structure of an

---

a) Current address: TCM, Cavendish Laboratory, Madingley Road, Cambridge CB3 0HE, UK
b) Current address: Department of Applied Physics, Yale University, New Haven, CT 06520, USA.

infinite Al monoatomic wire under compression. We also consider the effects of density functionals on cohesive energy and bond length of Al nanowires, and discuss the link from low dimensional 0D structure (dimer) to high dimensional 3D bulk structure of Al metal.

## 2. Method of computations

The wire calculations were performed by the plane wave ab initio pseudopotential method within the local density functional theory. The Hedin-Lundqvist (LDA) [22] form of exchange-correlation potential is used. Non-local norm-conserving pseudopotentials are created according to the prescription of Hamann et al [23]. The plane wave kinetic energy cutoff is 14.5 Ry which corresponds to about 90 plane waves per atom. During the self-consistent iterations, 21 special k-points in the Brillouin zone are used to calculate the screen potential and charge density. The chain configurations of Al atoms studied are the monoatomic linear chain, zigzag chain and di-atomic chain, as shown in Fig. 1. The supercell concept is used to generate a three-dimensional periodic tetragonal lattice with chains along the $z$ axis separated in the $x$ and $y$ directions. These separations between the chains are as wide as 8 Å, which is found to be large enough to isolate the chains, thereby effectively removing inter-chain interactions.

Another total energy package, CASTEP[29,30], has been used to cross-check our results and examine the density functional effect of the binding energy and bond length of Al nanowires. We performed systematic studies on Al nanostructures ranging from $Al_2$ dimer (0D), atomic chains (1D), monolayers (2D) to bulk Al (3D) with GGA-PW91[31] and LDA using an ultrasoft pseudopotential[32]. Special k points are generated by the Monkhorst-Pack (M-P) scheme[33]. The Γ k point (0,0,0) is used for dimer $Al_2$, (1x1x20) M-P k points are used for atomic chains, and M-P k point with spacing of $0.05Å^{-1}$ are generated for monolayers and bulk Al.

## 3. Results and discussions

We first perform a structural minimization for the free-standing linear chain of Al atoms, as shown in Fig. 1A. The calculated bond length of the Al chain is 2.40 Å, which is in good agreement with previous calculations of 2.38 Å [20]. This value is smaller than the calculated bond length of bulk fcc Al of 2.79 Å [24] due to a reduction in the coordination number going from the bulk system to the linear chain [25]. In agreement with the result in ref. [20], we did not find any tendency for the linear Al wire to dimerize, which is contrast to the linear gold wire [26]. Interestingly, such a difference is also observed between Al (110) and Au (110) surfaces. The former stabilizes at the (1x1) phase, while the later reconstructs into a (2x1) surface structure with dimerization [27].

The total energies (Ry/atom), Al-Al bond length, and bond angles of Al monatomic chain as a function of chain length are shown in Fig. 2. When we move alternate Al atoms slightly away from their original linear positions and then perform full relaxations, the Al chain reaches an optimized zigzag geometry, as shown in Fig. 1 (B), similar to the Au chain [18]. We found that the planar zigzag structure is more stable than the linear structure even when the wire is stretched, and the bond length increases in the zigzag geometry compared to the linear structure. This behavior is similar to that of Au [18] and Cu [28], but is different from that of K and Ca wires [28]. Sanchez-Portal et al [28] suggested that the differences between Au, Cu and K, Ca might be related to the presence of *d* bands at the Fermi level for the linear conformation of Au and Cu wires. However, there is no *d* band in the Al wire, only *s* and *p* bands in its electronic structure. Our results suggest that the zigzag geometry does not always occur in metal wires with *d* bands, but can also exist in metal wires without *d* bands such as Al.

In ref. [28], only Au is found to present two zigzag energy minima whilst K, Ca and Cu only stabilize in a single zigzag minimum as a function of the wire length. Surprisingly, Al wire is found to stabilize initially into an intermediate zigzag geometry with bond angle of $140^o$ and coordination number (CN) of 2, and subsequently into a global minimum of zigzag structure with equilateral triangles (CN=4, cf. di-atom chain in Fig. 1D). Al therefore represents a system without *d* bands that displays two zigzag energy minima in a metal nanowire. The origin of such anomalous behavior might be related to its detailed electronic structure. It can be seen in Fig. 2 that the bond length first decreases when it transits from being a stretched wire to the first zigzag energy minimum B, and then increases when the wire is contracted to the second zigzag energy minimum D (di-atom chain). The later can be understood in terms of the change in coordination number (CN) of the nanowire. In the first zigzag energy minimum B, CN = 2 and the bond length = 2.48 Å, which is slightly larger than the linear equilibrium bond length (2.40 Å). It can also be seen that for the zigzag structure B, the second neighbor distance (4.70 Å) decreases compared to that of the linear wire (4.80 Å). In order to maintain the optimum effective atomic coordination, the first-neighbor distance must therefore increase [28]. With the contracted wire approaching the second zigzag energy minimum D, the CN increases from 2 to 4, causing the bond length to increase to 2.68 Å.

The evolution of electronic structure of Al wire as a function of wire length should help us explain the stability of the zigzag geometry. The band structure and density of states (DOS) for the linear (A), first zigzag energy minimum (B), transition states (C) and second zigzag energy minimum (D) wires are presented in Fig. 3. In the linear wire (A), a π band (actually degenerated from originally two bands comprising $2p_2$ and $2p_3$ components) is seen near the Fermi level with high DOS, leading to the instability of the linear Al wire. For the zigzag rearrangement (B), these two degenerate bands near the Fermi level have separated and formed two electron channels connecting the valance and conduction bands. As the wire continues to be contracted, the bond angle decreases and the zigzag distortion increases; the bands composed of $2p_2$ and $2p_3$ components are distorted near the Fermi level and couple with the $2p_1$ component which goes down from the conduction band. The conduction bands cross the Fermi level increasing from two bands to three bands (see Fig. 3 transition structure C). The change of band structure and the addition of a third crossing band relate to the change in CN of the wire, and indicate the beginning of new chemical bond formation between second-nearest neighbors. When the chain reaches di-atom (D) geometry, these bands reshape with the $2p_1$ component moving above the Fermi level and into conduction band, while the $2p_2$ and $2p_3$ components form π bonds again and cross the Fermi level. The valence bands are pushed up towards the Fermi level near the (0,0,0.5) X k point, but are pulled down near the (0, 0, 0) Γ k point. The conduction band composed of the $2p_1$ component localized near the (0,0,0.5) k point forms an energy gap about 2 eV with one valence band also composed of the $2p_1$ component. These rearrangements of band structure lower the DOS near the Fermi level and thus lead to the stabilization of the di-atom structure.

Here we discuss the density functional effect (GGA and LDA) of the binding energy and bond length of Al nanowires. The calculated cohesive energies, bond lengths with GGA and LDA are listed in Table I. It can be seen from Table I that the cohesive energies and bond length obtained by GGA are generally smaller than those calculated from LDA. The effects of density functional (GGA or LDA) are more significant on cohesive energies than bond length. The fact that the cohesive energies obtained from GGA are close to experimental value indicates that the density functional is improved form LDA to GGA in Al. The choice of different density functional will also change the depth of energy valley of the zigzag di-atom chain (D) relative to the linear chain (A): GGA gives 0.743 eV/atom, while LDA gives 0.814 eV/atom.

To relate the atomic chain to bulk Al, we compare the energy and bond length of dimer $Al_2$, chains (A, B, and D), and three monolayers in (100), (110) and (111) orientations (as shown in Fig. 4), as well as bulk Al. The dimensions of these systems range from 0D to 3D. It is interesting to observe the energy and bond length change as a function of dimension. The cohesive energies and bond length of Al generally increase from 0D (dimer) to 3D (bulk). The (110) monolayer is an exception due to its larger geometrical relaxation and its significant change in size. The data for monolayers presented in Table I is obtained from optimized structures with fixed angles, i.e., the shape is fixed. Performing the full relaxation with both variable cell size and shape, the structures of monolayers in (100) and (111) orientations change a little, but the shape of (110) monolayer changes significantly and transforms into a low-energy structure with increased coordination number.

We note that the Al atomic chains considered here are all infinite chains. Although the structural and electronic information of such hypothetical structures provide basic knowledge of low-dimension nanostructures, the modeling of nanowires with more realistic features is essential to the understanding of experimental results with nanowires. With this in mind, we model a finite atomic chain suspended between two metal tips (the same element being used in both chain and tips), as shown in Fig 5. The geometry of the chain is full optimized while the atoms in tips are fixed. The bond lengths of Al-Al from one tip to another tip through atomic chain are shown in the Fig 5: the bond length of Al-Al near the tip is 2.539 Å, increasing to 2.637 Å and reaching largest value, 3.703 Å, in the middle of chain. The bond angle of the chain is 172.8°. The reason of this bond length arrangement can be explained as follows: The co-existing system of tip and chain is a low-dimension structure and with lower average CN compared with bulk Al, and thus the equilibrium bond length of Al-Al should be smaller than that in bulk Al. In this condition, the atomic chain is stretched. The Al atom in the beginning (or end) of the chain is connected with tip, the atoms in the tip trends to reduce its bond length due to dimensional effect, and thus this first Al atom in the chain is pulled close to tip. The same happen at the end of the Al chain connected with another tip. The total effect results in the larger gap (3.703 Å) between two atoms in the middle of the chain.

Adjusting the distance between the two tips can change the geometry of the atomic chain, and this can be compared with experiment. It is possible to simulate the structural evolution of atomic chains suspended by two tips by changing the position of tips from first principles calculations. Here we show an example for such "real" simulations at selected tip separations (bulk distance). A complete study of the dynamics of tip-suspended atomic chains will be useful to the understanding of experimental nanowires.

Table I. The cohensive energy, Ecoh (eV/atom), and bond length of Al atoms, $d$(Al-Al) (in Å) obtained by total energy package, CASTEP, with GGA-PW91 and LDA using an ultrasoft pseudopotential. Other available experimental and calculated data are also shown for comparison. (CN is coordination number).

| Systems | Dimension | Ecoh (eV/atom) | | ΔE(LDA-GGA) (eV/atom) | $d$(Al-Al) (Å) | | Δ$d$(LDA-GGA) (Å) |
|---|---|---|---|---|---|---|---|
| | | GGA | LDA | | GGA | LDA | |
| Dimer Al$_2$ | 0D (CN=1) | 0.695 | 0.941 | 0.246 | 2.596 | 2.596 | 0.000 |
| Linear chain (A) | 1D (CN=2) | 1.727 [1.87][a] | 2.017 | 0.290 | 2.368 [2.41][a] | 2.382 [2.40]* [2.38][d] | 0.014 |
| Zigzag chain (B) | 1D (CN=2) | 1.778 [1.92][a] | 2.070 | 0.292 | 2.472 (141.9°) [2.53][a] [139°][a] | 2.479 (141.4°) [2.48]* [140°]* | 0.007 |
| Zigzag chain (D) | 1D (CN=4) | 2.470 [2.65][a] | 2.831 | 0.361 | 2.492 (60°) [2.51][a] | 2.505 (60°) [2.68]* | 0.013 |
| (100) monolayer | 2D (CN=4) | 2.677 | 3.102 | 0.425 | 2.596 | 2.600 | 0.004 |
| (110) monolayer | 2D (CN=2) | 2.188 | 2.957 | 0.769 | 3.589, 2.343 | 2.744, 2.499 | 0.845, -0.156 |
| (111) monolayer | 2D (CN=6) | 2.903 | 3.338 | 0.435 | 2.623 | 2.633 | 0.010 |
| Bulk Al | 3D | 3.566 [3.67][a] [3.74][b] | 4.131 | 0.565 | 2.806 [2.86][a] [2.79][b] | 2.810 [2.79][c] [2.80][f] | 0.004 |
| Bulk Al (Expt.)[e] | | [3.39][e] | | | [2.86][e] [2.84][g] | | |

*This work using plane wave pseudopotential code with norm-conserving pseudopotentials.
[a]Ref[19]; [b]Ref[17]; [c]Ref[24]; [d]Ref[20]; [e]Cited from ref[19]; [f]Ref[34]; [g]Ref[35]

## 4. Conclusion

In conclusion, *ab initio* pseudopotential studies show that the Al nanowire, a non-*d*-band metal, displays two zigzag energy minima as a function of wire length, similar to transition metal Au but quite different from another transition metal Cu and simple metals K and Ca. The global minimum for the Al nanowire is a di-atom chain with zigzag geometry formed by equilateral triangles. The stability of this configuration is explained in terms of its electronic structure. The 2$p$ bands in the Al nanowire are shown to play a critical role in its stability. The effects of density functionals (GGA and LDA) on cohesive energy and bond length of Al nanostructures (dimer, chains, and monolayers) are also exmained. The link from low dimensional 0D structure (dimer) to high dimensional 3D bulk Al is estimated and the dimensional effect is discussed. We also show an example of tip-suspended finite atomic chain to fulfill the gap between hypothetical infinite chains and experimental finite chains.

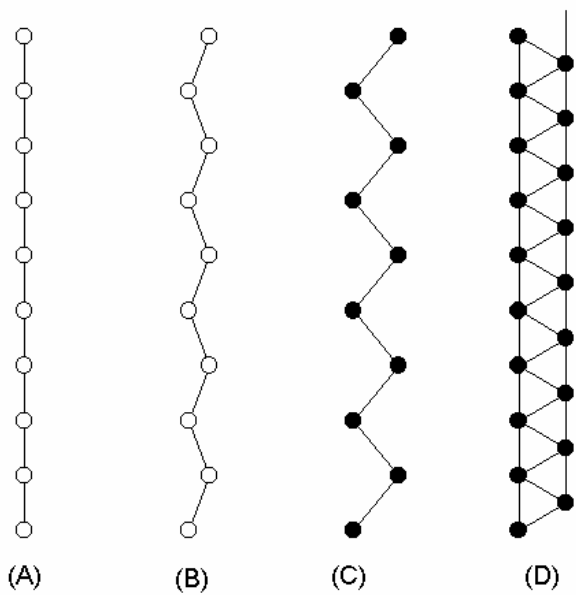

Fig 1. Geometrical configuration of Al chains: (A) linear chain, (B) first zigzag energy minimum structure with angle=140$^o$, (C) transition zigzag chain with angle=100$^o$, (D) global energy minimum structure, di-atomic chain (triangle) with angle=60$^o$.

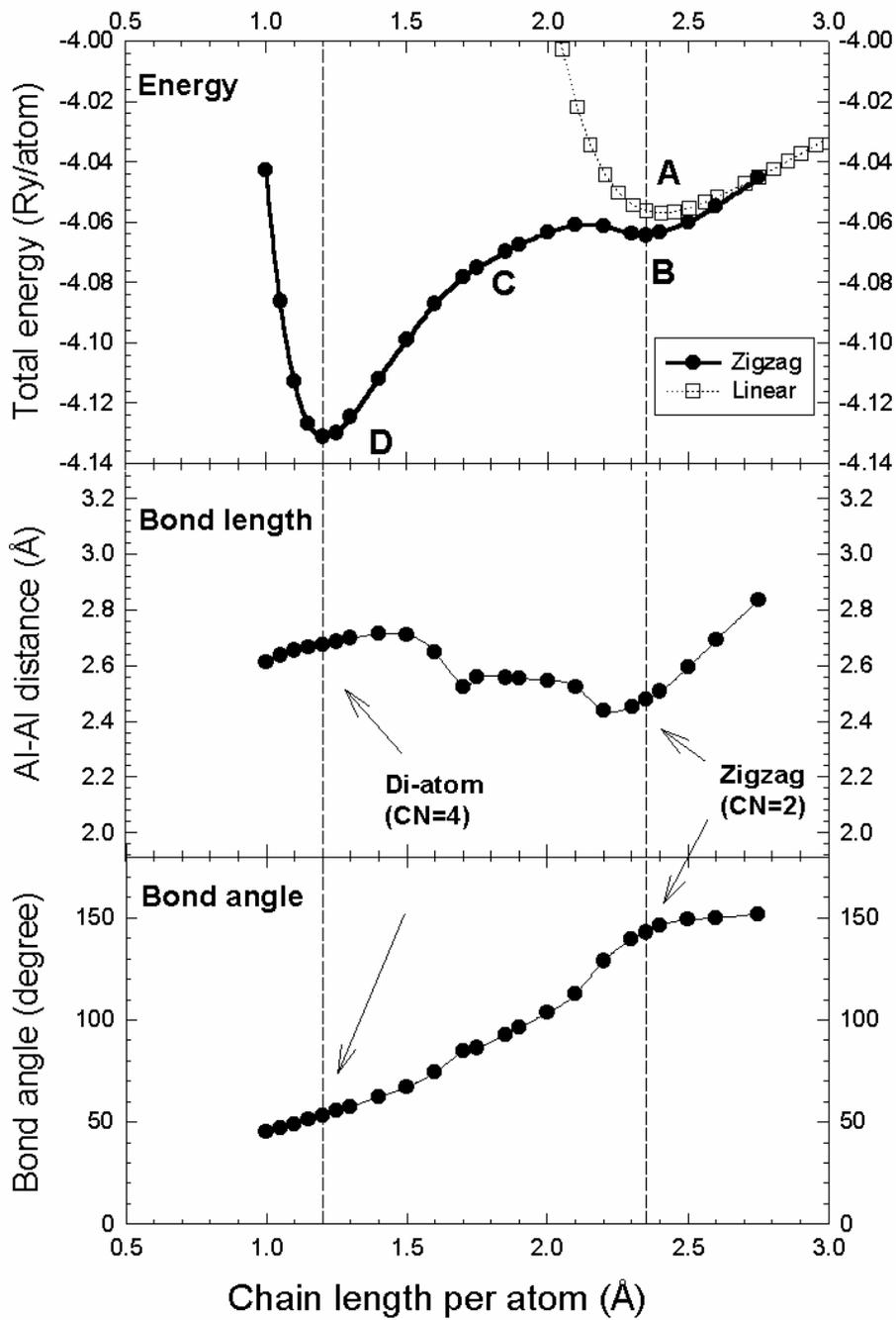

Fig 2. Total energies, bond length, and bond angle of monoatomic Al wire as a function of its length per atom. Total energy of linear chain is shown in dotted line with square data points.

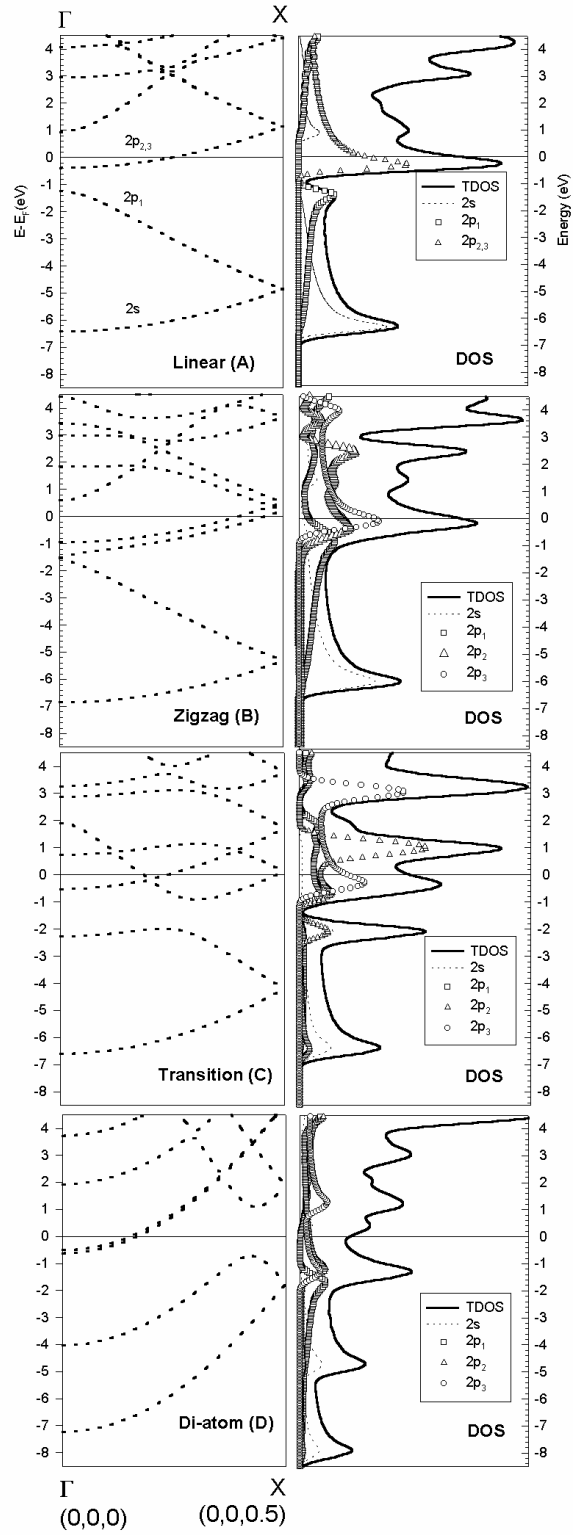

Fig 3. Band structures of typical geometries: the linear (A), $1^{st}$ zigzag (B) and transition zigzag (C) mono-atomic chains as well as that of equilibrium di-atomic chain (D).

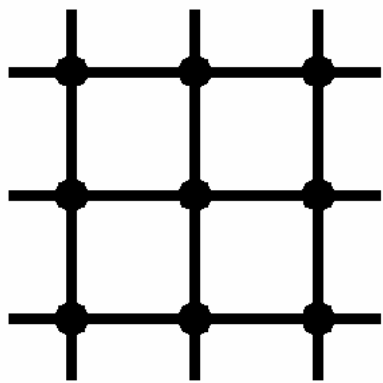

(a) (100)

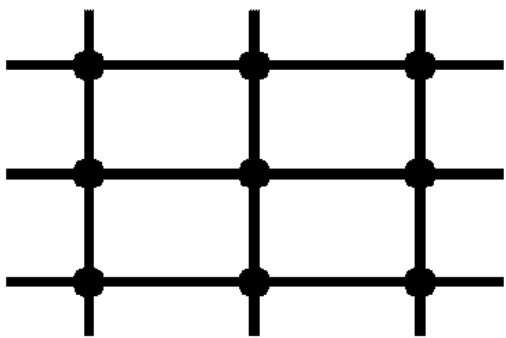

(b) (110)

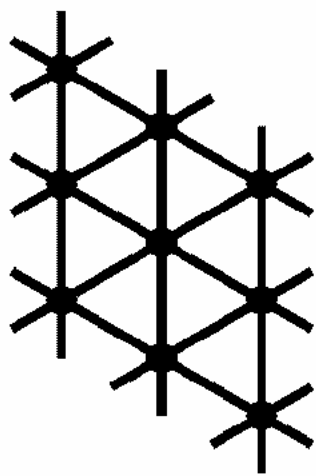

(c) (111)

Fig. 4. Topview of (a) Al(100), (b) Al(110), and (c) Al(111) monlayer structures.

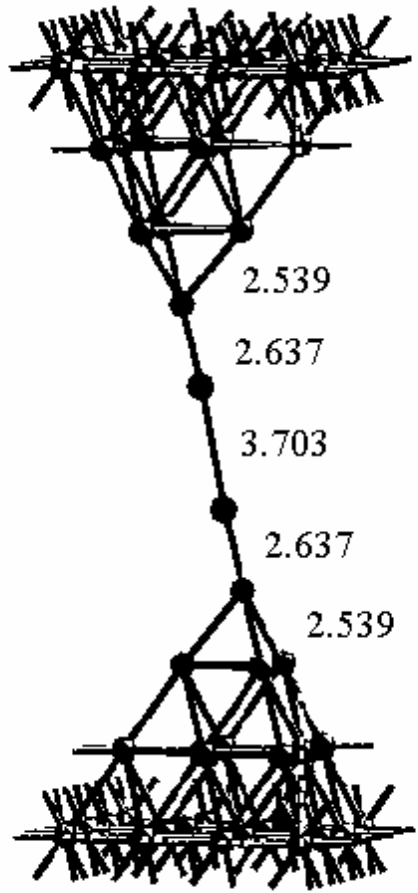

Fig. 5. Optimized structure of Al atomic chain (4 atoms) between two Al(111) tips. 4 atoms in Al atomic chain are allowed to fully relax, other atoms in two Al(111) tips are fixed during structural optimization. (unit is Å).